\documentclass[journal]{IEEEtran}
\usepackage{amsmath,amsfonts,amssymb}
\usepackage{algorithmic}
\usepackage{algorithm}
\usepackage{array}
\usepackage{textcomp}
\usepackage{stfloats}
\usepackage{url}
\usepackage{verbatim}
\usepackage{graphicx}
\usepackage{cite}
\usepackage{booktabs}
\usepackage{caption}
\usepackage{subfigure}
\usepackage{nomencl}
\usepackage{multirow}
\usepackage{threeparttable}
\usepackage{tabularx}
\usepackage{color}
\usepackage{courier}
\makeatletter
\newcommand{\removelatexerror}{\let\@latex@error\@gobble}
\makeatother
\removelatexerror

\begin{document}

\title{Improved Inner Approximation for Aggregating Power Flexibility in Active Distribution Networks and its Applications}

\author{Yilin~Wen,~\IEEEmembership{Graduate Student Member,~IEEE,}
        Zechun~Hu,~\IEEEmembership{Senior Member,~IEEE,}
        Jinhua~He, 
        and Yi~Guo,~\IEEEmembership{Member,~IEEE}
}



\maketitle

\begin{abstract}
  Concise and reliable modeling for aggregating power flexibility of distributed energy resources in active distribution networks (ADNs) is a crucial technique for coordinating transmission and distribution networks. Our recent research has successfully derived an explicit expression for the exact aggregation model (EAM) of power flexibility at the substation level under linearized distribution network constraints. The EAM, however, is impractical for decision-making purposes due to its exponential complexity. In this paper, we propose an inner approximation method for aggregating flexibility in ADNs that utilizes the properties of the EAM to improve performance. Specifically, the geometric prototype of the inner approximation model is defined according to a subset of the coefficient vector set of the EAM, which enhances the accuracy. On the other hand, the computation efficiency of the inner approximation is also significantly improved by exploiting the regularity of coefficient vectors in the EAM in the parameter calculation process. The inner approximated flexibility model of ADNs is further incorporated into the security-constrained unit commitment problem as an application. Numerical simulations verify the effectiveness of the proposed method.
\end{abstract}


\section{Introduction}
\IEEEPARstart{V}{arious} distributed energy resources (DERs) are increasingly connected to distribution networks, bringing about a non-negligible impact on the power system's operation. Typical DERs, such as electric vehicles (EVs), distributed energy storage systems (DESSs), and rooftop photovoltaics (PVs), offer flexibility in their power schedules  \cite{stanojev2022multiple}. Their integration enables distribution networks to adjust their power actively. Active distribution networks (ADNs) are forming, which will become a critical dispatchable resource in future power systems where more and more traditional generators will be replaced by renewable energy. Since distribution and transmission networks are interconnected and adjustments in one can affect the other, it is imperative for power system operators to implement coordinated dispatch of transmission and distribution networks to ensure the safety and economics of the entire system \cite{lind2019transmission,givisiez2020review}.

Exploiting the power flexibility of ADNs for power system operation scheduling poses significant technical challenges. The primary issue is how to effectively model ADNs in the decision-making at the transmission level. Centralized computing that considers all flexible ranges of DERs and distribution network constraints by the dispatch center is impractical due to the curse of dimensionality. To this end, many researchers are exploring practical and viable methods to address DERs and ADNs at the transmission level \cite{kargarianSystemSystemsBased2014, yuanHierarchicalCoordinationTSODSO2017, yinDistributionallyRobustDecentralized2022, zhaiDistributionallyRobustJoint2022, liuAccurateCompactModel2021,liuIntegratingHighDERPenetrated2021,patigFastMappingFlexibility2022,kalantar-neyestanakiCharacterizingReserveProvision2020,silvaEstimatingActiveReactive2018,wenAggregateFeasibleRegion2022,wenAggregateTemporallyCoupled2022,chenAggregatePowerFlexibility2020,yanDistributedCoordinationCharging2023, chenLeveragingTwoStageAdaptive2021,cuiNetworkCognizantTimeCoupledAggregate2021,mullerAggregationEnergeticFlexibility2015,mullerAggregationDisaggregationEnergetic2019,haoAggregateFlexibilityThermostatically2015, zhaoGeometricApproachAggregate2017,wangAggregateFlexibilityVirtual2021,yiAggregateOperationModel2021,zhangCharacterizingTemporalCoupledFeasible2022}.

Decomposition techniques have been proposed to handle the massive DERs and ADNs in the coordinated dispatch of transmission and distribution networks\cite{kargarianSystemSystemsBased2014, yuanHierarchicalCoordinationTSODSO2017, yinDistributionallyRobustDecentralized2022, zhaiDistributionallyRobustJoint2022}. These methods involve iterative computation and multiple information exchanges, which require high computation and communication capabilities. Alternatively, a more promising approach is to concisely model the aggregated power flexibility at the transmission-distribution interface (physically, the substation). Through this method, the decision-making at the transmission level does not need to take care of details in the ADN but only needs to consider the aggregated flexibility model at the substation level. The scheduled power profile of the substation is then disaggregated to each DER in the ADN. This framework does not require iterations, thus reducing the need for computation and communication resources and lowering the implementation threshold.

The modeling of aggregated power flexibility has been the subject of much recent research. The Fourier-Motzkin Elimination method has been used in \cite{liuAccurateCompactModel2021,liuIntegratingHighDERPenetrated2021,patigFastMappingFlexibility2022} to generate the constraints on the power at the transmission-distribution interface, accounting for linear distribution network constraints. A linear scenario-based robust optimization approach is also applied in \cite{kalantar-neyestanakiCharacterizingReserveProvision2020} to form the adjustable range of the ADN's active and reactive power range. Reference \cite{silvaEstimatingActiveReactive2018} calculates the approximated adjustable power range of the ADN considering nonlinear power flow constraints via interior point method. However, all these methods mentioned above face the challenge of high computational complexity when taking into account the coupling among the power across multiple time slots. The temporal coupling of power flexibility is common, such as the energy constraints of EV and DESS. Failing to address this coupling can lead to major discrepancies between the calculated flexible power range and the realizable range.

Several studies have proposed aggregation methods that can handle the temporal coupling of power flexibility. In our prior work \cite{wenAggregateFeasibleRegion2022}, we derived the explicit expression for the exact aggregation model (EAM) of multiple DERs. And in \cite{wenAggregateTemporallyCoupled2022}, we demonstrated that the EAM retains its form when considering linearized distribution network constraints. It has been observed that the number of constraints in the EAM increases exponentially with the number of time slots, making it challenging to implement in practical applications. A series of approximation models are thus proposed to reduce the complexity at the expense of accuracy. The approximation models in \cite{wenAggregateFeasibleRegion2022} and \cite{wenAggregateTemporallyCoupled2022} are all outer approximations, which means that there may exist some power profiles in these models that cannot be disaggregated to each DER.

Many researchers are exploring inner approximations to increase the reliability of the aggregate flexibility model. Most of the existing inner approximation methods can be summarized into two steps: first, defining the geometric prototype of the approximation model, and second, calculating the parameters in the prototype. Previous literature studied many prototypes such as box \cite{chenAggregatePowerFlexibility2020,yanDistributedCoordinationCharging2023, chenLeveragingTwoStageAdaptive2021}, ellipse \cite{chenLeveragingTwoStageAdaptive2021, cuiNetworkCognizantTimeCoupledAggregate2021}, zonotope \cite{mullerAggregationEnergeticFlexibility2015,mullerAggregationDisaggregationEnergetic2019}, virtual battery model \cite{haoAggregateFlexibilityThermostatically2015, zhaoGeometricApproachAggregate2017}, power-energy boundary model \cite{wangAggregateFlexibilityVirtual2021}, and homothetic polytope \cite{yiAggregateOperationModel2021}. Various parameter calculation methods have been designed based on the characteristics of each prototype. Reference \cite{zhangCharacterizingTemporalCoupledFeasible2022} does not pre-set any prototype but forms the inner approximation model by continuously adding feasible cuts. In the current inner approximation methods, the prototype formulation (or initial parameters in the case of \cite{zhangCharacterizingTemporalCoupledFeasible2022}) is often based on empirical criteria. As a result, there is a risk of over-conservatism in the resulting models. Meanwhile, the computational complexity of some previous parameter calculation methods limits their practicality. For instance, reference \cite{cuiNetworkCognizantTimeCoupledAggregate2021} only tested up to 4 time slots in the simulation, \cite{wangAggregateFlexibilityVirtual2021} tested up to 8, and \cite{chenLeveragingTwoStageAdaptive2021} tested up to 15 for the box prototype and 6 for the ellipse prototype. However, in practical applications such as unit commitment (UC), at least 24 slots need to be considered. Hence, if ADNs are to be incorporated into UC \cite{kargarianSystemSystemsBased2014}, these inner approximation models may be inapplicable.

To summarize, the high complexity of exact flexibility aggregation necessitates approximation methods, especially inner approximation methods with higher reliability, for the coordination of transmission and distribution networks. However, all inner approximation methods the authors have seen so far were proposed without knowing the explicit expression of the EAM. The accuracy and computational efficiency should be improved to make it applicable in practice. Therefore, this paper proposes a novel inner approximation method that leverages properties of the EAM in both the prototype selection and the parameter calculation processes, improving the accuracy and reducing the complexity. Furthermore, the proposed inner approximation method for aggregating power flexibility of ADNs is applied to the security-constrained unit commitment (SCUC) in the numerical tests, demonstrating the practicality.

The rest of this paper is organized as follows. Section II describes the proposed inner approximation method without considering distribution network constraints. In Section III, we introduce a double inner approximation framework to deal with a large number of DERs and network constraints in the ADN. Numerical simulations are carried out in Section IV. Finally, Section V concludes this paper.

\section{Inner Approximation Method based on the Exact Aggregation Model}\label{sect:innerapprox}
This section introduces the inner approximation method for aggregating the DER flexibility without distribution network constraints, especially how the properties of the EAM are utilized in the approximation process. Incorporating network constraints will be discussed in the next section. We start the discussion with a generalized individual flexibility model of DERs and its exact aggregation.
\subsection{Generalized Individual Flexibility Model of DERs and the Exact Aggregation Model}
Let us discretize the time horizon into $T$ slots with the length of $\Delta T$. Denote by $t$ the indices of time slots and define $[T] \triangleq \{1,2,\cdots,T\}$. Then, the individual flexibility of typical types of DERs such as EVs, DESSs, and PVs can be uniformly described as the following power-energy boundary model:\begin{subequations}\label{eq:powerenergybound}
  \begin{align}
    &{\underline p _{n,t}} \le {p_{n,t}} \le {\overline p _{n,t}},\forall t \in [T],\forall n \in \mathcal N,\label{eq:powerbound}\\
    &{\underline e _{n,t}} \le \Delta T\sum\nolimits_{\tau  = 1}^t {{p_{n,\tau }}}  \le {\overline e _{n,t}},\forall t \in [T],\forall n \in \mathcal N,\label{eq:energybound}
  \end{align}
\end{subequations}
where, $n/\mathcal N$ denotes the indices/set of DERs, $|\mathcal N|=N$; $p_{n,t}$ denotes the power of DER $n$ at time $t$; ${\underline p _{n,t}}$, ${\overline p _{n,t}}$, ${\underline e_{n,t}}$, and ${\overline e_{n,t}}$ are the power and energy boundary parameters. Equations \eqref{eq:powerbound} and \eqref{eq:energybound} constrain the power and accumulated energy consumption of each DER, respectively. These boundary parameters are influenced by the properties of different DERs. For example, the power of an EV is adjustable during its connecting period, but it should be charged to an expected level before departure. Similiarly, DESS should restore its energy to the initial value at the end of the time horizon. The calculation details of these parameters for each specific DER can be found in \cite{xuHierarchicalCoordinationHeterogeneous2016}. 

The EAM refers to a set of constraints ensuring that the aggregated power $P_t \triangleq \sum\nolimits_{n\in \mathcal N}p_{n,t}, \forall t \in [T]$ can be disaggregated to each DER if and only if $P_t(\forall t \in [T])$ satisfies these constraints. According to our previous derivations in \cite{wenAggregateFeasibleRegion2022}, these constraints are
\begin{equation}\label{eq:EAM_vec}
  \underline {\phi} _{\mathbf{u}} \le \mathbf{u}^{\top}\mathbf{P}  \le \overline {\phi} _{\mathbf{u}} ,\forall \mathbf{u} \in \mathbb B^T, \mathbf{u} \neq \mathbf{0},
\end{equation}
where, $\mathbf P$ denotes the $T\times 1$ vector composed of $P_t(\forall t \in [T])$, $(\cdot)^{\top}$ denotes the vector transposition, $\mathbb B \triangleq \{0,1\}$ so that $\mathbb B^T$ is the set of all $T\times 1$ vectors whose components must be $1$ or $0$, $\mathbf u$ denotes the coefficient vector, $\underline {\phi} _{{\mathbf u}}$ and $\overline {\phi} _{{\mathbf u}}$ are parameters generated from $\Delta T$, ${\underline p _{n,t}}$, ${\overline p _{n,t}}$, ${\underline e_{n,t}}$, and ${\overline e_{n,t}}$ (see \cite{wenAggregateFeasibleRegion2022} for details). It is observed from \eqref{eq:EAM_vec} that power variables $P_t$ of any non-empty subset in the time horizon are coupled. This coupling leads to the exponential number of constraints in the EAM with respect to the number of time slots $T$. Hence, approximation models are necessary to make the aggregated flexibility model practical.
\subsection{Inner Approximation Method}
An inner approximation is a feasible region that is fully contained within the EAM and can be described with lower complexity than the EAM. It guarantees that all the inside power profiles can be disaggregated into each DER. The fundamental idea of many existing inner approximation methods can be summarized as follows: First, define a geometric prototype of the approximate region with some undetermined parameters. Then, calculate these parameters to make the approximate region as large as possible while ensuring that it remains fully contained within the EAM. The proposed method is also born on this basic framework.
\subsubsection{Define the Geometric Prototype for the Approxmation}\;

Some prototypes based on experience and intuition have been studied in the literature. For example, references \cite{chenAggregatePowerFlexibility2020} and \cite{chenLeveragingTwoStageAdaptive2021} use the power boundary prototype
\begin{equation}\label{eq:approx_PB}
  {\underline P_{t}} \le {P_{t}} \le {\overline P_{t}},\forall t \in [T]
\end{equation}
with $2T$ constraints, where parameters ${\underline P_{t}}$ and ${\overline P_{t}}$ are to be determined.
In \cite{wangAggregateFlexibilityVirtual2021}, the power-energy boundary prototype
\begin{subequations}\label{eq:approx_PEB}
  \begin{align}
    &{\underline P_{t}} \le {P_{t}} \le {\overline P_{t}},\forall 1\leq t \leq T,\label{eq:PEB_PB}\\
    &{\underline E_{t}}/ \Delta T \le\sum\nolimits_{\tau  = 1}^t {{P_{\tau }}}  \le {\overline E_{t}}/ \Delta T,\forall 2\leq t \leq T \label{eq:PEB_EB}
  \end{align}
\end{subequations}
with $4T-2$ constraints is studied, where parameters ${\underline P_{t}}$, ${\overline P_{t}}$, ${\underline E_{t}}$ and ${\overline E_{t}}$ are to be determined.

A closer look reveals that the sets of coefficient vectors in the above two prototypes are actually subsets of $\mathbb B^T$. In principle, any shape can be used as the prototype, but selecting the prototype with a subset of $\mathbb B^T$ offers two advantages: i) it makes the approximate model similar in shape to the EAM, thereby improving accuacy; ii) the initial parameter values are easy to define, which enhances computation efficiency. However, the power boundary and power-energy boundary prototypes correspond to very small subsets of $\mathbb B^T$. Higher accuracy may be reached by defining the prototype with a larger subset of $\mathbb B^T$, such as the energy-change boundary prototype\footnote{It seems that the summation term in \eqref{eq:approx_ECB} needs to be multiplied by $\Delta T$ to represent the energy change, but here $\Delta T$ is equivalently incorporated into parameters $\underline {{\phi}} _{{t_1},{t_2}}$ and $\overline {{\phi}} _{{t_1},{t_2}}$ for simplicity.}:
\begin{equation}\label{eq:approx_ECB}
  \underline {{\phi}} _{{t_1},{t_2}}  \le \sum\nolimits_{t = {t_1}}^{{t_2}} {{P_t}}  \le \overline {{\phi}} _{{t_1},{t_2}} ,\forall 1\leq{t_1} \le {t_2} \le T,
\end{equation}
with $T(T+1)$ constraints, where parameters $\underline {{\phi}} _{{t_1},{t_2}}$ and $\overline {{\phi}} _{{t_1},{t_2}}$ are to be determined. This prototype constrains the energy change between every two slots, which has been studied as an outer approximation in \cite{wenAggregateFeasibleRegion2022} and achieves good accuracy.

In the following text, for convenience, the EAM \eqref{eq:EAM_vec} is compactly represented as
\begin{equation}\label{eq:EAM_compact}
  {\Psi _{{\text{ext}}}} \triangleq \left\{ {\mathbf{P}}|{{\mathbf{A}}_{{\text{ext}}}}{\mathbf{P}} \le {{\mathbf{b}}_{{\text{ext}}}}\right\},
\end{equation}
where, $\mathbf{A}_{\text{ext}}$ and $\mathbf{b}_{\text{ext}}$ are formed based on \eqref{eq:EAM_vec}. Approximate models are uniformly represented as
\begin{equation}\label{eq:Approx_compact}
  {\Psi_{\text{apx}}} \triangleq \left\{ {\mathbf{P}}|{{\mathbf{A}}_{{\text{apx}}}}{\mathbf{P}} \le {{\mathbf{b}}_{{\text{apx}}}}\right\},
\end{equation}
where, $\mathbf{A}_{\text{apx}}$ is a fixed matrix of coefficient vectors and $\mathbf{b}_{\text{apx}}$ is the parameter vector to be determined. The number of rows of $\mathbf{A}_{\text{apx}}$ is denoted by $N_c$, so the power boundary, power-energy boundary, and energy-change boundary prototypes correspond to $N_c=2T$, $N_c=4T-2$, and $N_c=T(T+1)$, respectively.
\subsubsection{Calculate Parameters in the Prototype}\;

This paper proposes an iterative method to determine the parameter $\mathbf{b}_{\text{apx}}$. First, initialize $\mathbf{b}_{\text{apx}}$, namely $\mathbf{b}^{(0)}_{\text{apx}}$. Since the coefficient vector set in $\Psi^{(0)}_{\text{apx}}$ is a subset of $\mathbb B^T$, $\mathbf{b}^{(0)}_{\text{apx}}$ can be directly extracted from its corresponding part in $\mathbf{b}_{\text{ext}}$ of $\Psi_{\text{ext}}$ using the parameter calculation formula\footnote{This formula calculates the components of $\mathbf{b}_{\text{ext}}$ one by one through simple comparison and summation operations. Here, we only need to calculate the $N_c$ components of $\mathbf{b}_{\text{ext}}$ as $\mathbf{b}^{(0)}_{\text{apx}}$, rather than calculating all components, so there is no need to worry about the high complexity of $\Psi_{\text{ext}}$.} in \cite{wenAggregateFeasibleRegion2022}. The initial approximation model $\Psi^{(0)}_{\text{apx}}$ thus obtained is an outer approximation of $\Psi_{\text{ext}}$, and then $\mathbf{b}_{\text{apx}}$ is modified iteratively to shrink $\Psi_{\text{apx}}$ to the interior of $\Psi_{\text{ext}}$. The specific process for modifying $\mathbf{b}_{\text{apx}}$ is introduced as follows.

Let the superscript $(k)$ denote the $k$-th iteration. Before the iteration converges, $\Psi^{(k-1)}_{\text{apx}}\nsubseteq \Psi_{\text{ext}}$. Notice that
\begin{align}
  \Psi^{(k-1)}_{\text{apx}}\nsubseteq \Psi_{\text{ext}} \Leftrightarrow& \exists \text{ a point }\mathbf{P}_1 \in \Psi^{(k-1)}_{\text{apx}},\mathbf{P}_1 \notin  \Psi_{\text{ext}},\nonumber\\
  \Leftrightarrow& \exists \mathbf{P}_1 \in \Psi^{(k-1)}_{\text{apx}} \text{ and a row } j \text{ of }\mathbf{A}_{\text{ext}}, \text{ such} \nonumber\\
  &\text{that }\mathbf{a}_{\text{ext},j}\mathbf{P}_1> b_{\text{ext},j} \;\; (\textbf{Prop.1}), \nonumber
\end{align}
where, $\mathbf{a}_{\text{ext},j}$ is the $j$-th row of $\mathbf{A}_{\text{ext}}$ and $b_{\text{ext},j}$ is the $j$-th component of $\mathbf b_{\text{ext}}$. According to Equation \eqref{eq:EAM_vec}, the row vector $\mathbf{a}_{\text{ext},j}$ can only be $\mathbf u^{\top}$ or $-\mathbf u^{\top}$ $(\forall \mathbf u \in \mathbb B^T)$ and $b_{\text{ext},j} = \max\nolimits_{{{\mathbf{P}}_0} \in {\Psi _{\text{ext}}}} \mathbf{a}_{\text{ext},j}{\mathbf{P}}_0$. Therefore, \textbf{Prop.1} is equivalent to: $\exists \mathbf u \in \mathbb B^T$ such that inequality \eqref{eq:error_pos} or \eqref{eq:error_neg} holds.
\begin{subequations}\label{eq:def_error}
  \begin{align}
    \max \limits_{{{\mathbf{P}}_1} \in \Psi^{(k-1)}_{\text{apx}}} {{\mathbf{u}}^{\top}}{{\mathbf{P}}_1} - \max \limits_{{{{\mathbf{P}}_0} \in {\Psi _{\text{ext}}}}} {{\mathbf{u}}^{\top}}{{\mathbf{P}}_0} &> 0 \label{eq:error_pos}\\
    \min \limits_{{{{\mathbf{P}}_0} \in {\Psi _{\text{ext}}}}} {{\mathbf{u}}^{\top}}{{\mathbf{P}}_0} -\min \limits_{{{\mathbf{P}}_1} \in \Psi^{(k-1)}_{\text{apx}}} {{\mathbf{u}}^{\top}}{{\mathbf{P}}_1} &> 0 \label{eq:error_neg}
  \end{align}
\end{subequations}

The left sides of Equations \eqref{eq:error_pos} and \eqref{eq:error_neg} reflect the gaps between $\Psi^{(k-1)}_{\text{apx}}$ and $\Psi_{\text{ext}}$ in the positive and negative directions of ${\mathbf{u}}$, respectively. If we can find the ${\mathbf{u}}$ that maximizes this gap, the corresponding $\mathbf P_1$ and $\mathbf P_0$ can be used to update $\mathbf b_{\text{apx}}$ so that shrinking $\Psi^{(k-1)}_{\text{apx}}$ to the inside of $\Psi_{\text{ext}}$ in the positive and negative direction of ${\mathbf{u}}$. To this end, solve problem \eqref{eq:origin_bi_pos} for the positive direction and \eqref{eq:origin_bi_neg} for the negative direction.
\begin{subequations}\label{eq:origin_bi_pos}
  \begin{align}
    d^{(k)}_{{\text{pos}}} \triangleq \max\limits_{\mathbf{u}} \left( {\max\limits_{{{\mathbf{P}}_1}} {{\mathbf{u}}^{\top}}{{\mathbf{P}}_1} - \max\limits_{{{\mathbf{P}}_0}} {{\mathbf{u}}^{\top}}{{\mathbf{P}}_0}} \right)& \label{eq:origin_bi_obj}\\
    \text{s.t.  }{\mathbf{u}} \in {\mathbb B ^T},{{\mathbf{P}}_1} \in \Psi^{(k-1)}_{\text{apx}},{{\mathbf{P}}_0} \in {\Psi _{{\text{ext}}}},& \label{eq:origin_bi_cons}
  \end{align}
\end{subequations}
\begin{equation}\label{eq:origin_bi_neg}
    d^{(k)}_{{\text{neg}}} \triangleq \max\limits_{\mathbf{u}} \left( {\min\limits_{{{\mathbf{P}}_0}} {{\mathbf{u}}^{\top}}{{\mathbf{P}}_0} - \min\limits_{{{\mathbf{P}}_1}} {{\mathbf{u}}^{\top}}{{\mathbf{P}}_1}} \right)  \text{  s.t.  } \eqref{eq:origin_bi_cons},
\end{equation}
where, $d^{(k)}_{{\text{pos}}}/d^{(k)}_{{\text{neg}}} $ denotes the maximum gap in all the positive/negative directions.

The above two problems are bi-level optimization problems and the objective functions contain bilinear terms, so they cannot be solved directly. We transform problems \eqref{eq:origin_bi_pos} and \eqref{eq:origin_bi_neg} into equivalent Mixed-Integer Linear Programming (MILP) problems which can be efficiently solved. Take problem \eqref{eq:origin_bi_pos} as an example to introduce the transformation process in detail.

First, rewrite \eqref{eq:origin_bi_pos} as an equivalent minimization problem:
\begin{equation}\label{eq:trans_to_min}
    \min\limits_{\mathbf{u},\mathbf{P}_1} \left({ {-{\mathbf{u}}^{\top}}{{\mathbf{P}}_1} + \max\limits_{{{\mathbf{P}}_0}} {{\mathbf{u}}^{\top}}{{\mathbf{P}}_0}} \right)  \text{  s.t.  } \eqref{eq:origin_bi_cons}.
\end{equation}
The inner maximum problem can be further transformed into a minimum problem using the duality theorem. However, direct transformation leads to unacceptable complexity since the explicit expression of $\Psi _{\text{ext}}$ includes $2(2^T-1)$ constraints. Hence, we use the equivalent form:
\begin{equation}\label{eq:EAM_eqv}
  {\Psi _{{\text{ext}}}} \Leftrightarrow \left\{ {{{\mathbf{P}}_0}|{{\mathbf{P}}_0} = \sum\nolimits_{n \in {{\mathcal N}}} {{{\mathbf{p}}_n}} ,\eqref{eq:powerenergybound}} \right\}, 
\end{equation}
where, $\mathbf{p}_n$ is the $T\times 1$ vector composed by $p_{n,t}, \forall t\in [T]$. Arrange $\mathbf{p}_n,\forall n\in \mathcal N$ into an $NT\times 1$ vector $\mathbf x$, i.e. $\mathbf x = [\mathbf{p}_1^{\top},\mathbf{p}_2^{\top},\cdots,\mathbf{p}_N^{\top}]^{\top}$, then rewrite \eqref{eq:EAM_eqv} as the matrix form:
\begin{equation}\label{eq:EAM_eqv_matrix}
  {\Psi _{{\text{ext}}}} \Leftrightarrow \left\{ {{{\mathbf{P}}_0} | {{\mathbf{Cx}} - {{\mathbf{P}}_0} = {\mathbf{0}},{\mathbf{Dx}} \le {\mathbf{f}}} } \right\}, 
\end{equation}
where, $\mathbf C$ is a $T\times NT$ coefficient matrix generated by ${{\mathbf{P}}_0} = \sum\nolimits_{i \in {{\mathcal N}}} {{{\mathbf{p}}_i}}$, $\mathbf D$ is a $4NT\times NT$ sparse matrix and $\mathbf f$ is a $4NT \times 1$ vector generated from \eqref{eq:powerenergybound}. According to the duality theorem, problem \eqref{eq:trans_to_min} can be equivalently transformed into
\begin{subequations}\label{eq:trans_to_dual}
\begin{align}
  &\min  - {{\mathbf{u}}^{\top}}{{\mathbf{P}}_1} + {{\boldsymbol{\lambda }}^{\top}}{\mathbf{f}} \label{eq:trans_to_dual_obj}\\
  &\text{s.t.  }{\mathbf{u}} \in {\mathbb B ^T},{{\mathbf{P}}_1},{\boldsymbol{\omega }} \in {\mathbb{R}^T}, {\boldsymbol{\lambda }} \in {\mathbb{R}^{4NT}},\nonumber \\
  &{{\mathbf{A}}_{{\text{apx}}}}{{\mathbf{P}}_1} \le {{\mathbf{b}}^{(k-1)}_{{\text{apx}}}},\label{eq:trans_to_dual1}\\
  &{\mathbf{u}} + {\boldsymbol{\omega }} = \mathbf{0},\label{eq:trans_to_dual2}\\
  &{{\mathbf{C}}^{\top}}{\boldsymbol{\omega }} + {{\mathbf{D}}^{\top}}{\boldsymbol{\lambda }} = \mathbf{0},\label{eq:trans_to_dual3}\\
  &{\boldsymbol{\lambda }} \ge \mathbf{0},\label{eq:trans_to_dual4}
\end{align}
\end{subequations}
where, $\boldsymbol{\omega }$ and $\boldsymbol{\lambda }$ are dual variables, $\mathbb{R}^m$ denotes the $m$-dimensional Euclidean space. 

There is still a bilinear term in the objective. Fortunately, since $\mathbf u$ is a binary vector, problem \eqref{eq:trans_to_dual} can be equivalently transformed into \eqref{eq:trans_to_final} by introducing an auxiliary variable.
\begin{subequations}\label{eq:trans_to_final}
  \begin{align}
    &\min  {{\mathbf{1}}^{\top}}\mathbf y + {{\boldsymbol{\lambda }}^{\top}}{\mathbf{f}}\label{eq:trans_to_final_obj}\\
    &\text{s.t.  }{\mathbf{u}} \in {\mathbb B ^T},{{\mathbf{P}}_1},{\boldsymbol{\omega }}, \mathbf y \in {\mathbb{R}^T},{\boldsymbol{\lambda }} \in {\mathbb{R}^{4NT}},\nonumber \\
    &\eqref{eq:trans_to_dual1} - \eqref{eq:trans_to_dual4}, \nonumber \\
    &-M\mathbf u\leq \mathbf y \leq M\mathbf u\label{eq:trans_to_final_1},\\
    & - M(\mathbf 1 - {\mathbf{u}}) \le \mathbf y +\mathbf P_1 \le M(\mathbf 1 - {\mathbf{u}})\label{eq:trans_to_final_2},
    \end{align}
\end{subequations}
where, $\mathbf y$ is the introduced auxiliary variable, $M$ is a large-enough constant. The above problem is an MILP with only $T$ binary variables. The number of binary variables \textbf{does not change} with the prototype selection of $\Psi_{\text{apx}}$. The negative direction problem \eqref{eq:origin_bi_neg} can be similarly transformed into:
\begin{subequations}\label{eq:trans_to_final_neg}
  \begin{align}
    &\min  {{\mathbf{1}}^{\top}}\mathbf y + {{\boldsymbol{\lambda }}^{\top}}{\mathbf{f}}\label{eq:trans_to_final_neg_obj}\\
    &\text{s.t.  }{\mathbf{u}} \in {\mathbb B ^T},{{\mathbf{P}}_1},{\boldsymbol{\omega }}, \mathbf y \in {\mathbb{R}^T},{\boldsymbol{\lambda }} \in {\mathbb{R}^{4NT}},\nonumber \\
    &\eqref{eq:trans_to_dual1},\eqref{eq:trans_to_dual3},\eqref{eq:trans_to_dual4},\eqref{eq:trans_to_final_1}, \nonumber \\
    &-{\mathbf{u}} + {\boldsymbol{\omega }} = \mathbf{0},\label{eq:trans_to_final_neg_1}\\
    & - M(\mathbf 1 - {\mathbf{u}}) \le \mathbf y -\mathbf P_1 \le M(\mathbf 1 - {\mathbf{u}}).\label{eq:trans_to_final_neg_2}
    \end{align}
\end{subequations}

Solving problems \eqref{eq:trans_to_final} and \eqref{eq:trans_to_final_neg} yields the optimal $\mathbf{P}_1$, denoted by $\mathbf{P}^{(k)}_1$, in the positive and negative directions, respectively. However, the primal variable $\mathbf P_0$ is lost during the transformation. Hence, solve the following problem to find $\mathbf{P}^{(k)}_0$---the closest point to $\mathbf{P}^{(k)}_1$ in $\Psi _{{\text{ext}}}$:
\begin{equation}\label{eq:find_P0}
  {\mathbf{P}^{(k)}_0} = \mathop {\arg \min }\limits_{{{\mathbf{P}}_0}} \left\| {{{\mathbf{P}}_0} - {\mathbf{P}^{(k)}_1}} \right\|,{\text{s.t.  }}{\mathbf{Cx}} - {{\mathbf{P}}_0} = {\mathbf{0}},{\mathbf{Dx}} \le {\mathbf{f}},
\end{equation}
where, $\|\cdot\|$ calculates the $\ell_2 $-norm in the $T$-dimensional Euclidean space.

Next, update ${\mathbf{b}}_{\text{apx}}$ according to $\mathbf{P}^{(k)}_0$ and $\mathbf{P}^{(k)}_1$. We start by defining the set of active constraints for $\mathbf{P}^{(k)}_1$ in $\Psi^{(k-1)}_{\text{apx}}$, i.e., 
\begin{equation}\label{eq:set_of_active}
  {\mathcal{J}^{(k)}} \triangleq \left\{ {j\left |{{\mathbf{a}}_{\text{apx},j}}{\mathbf{P}}_1^{(k)} = b_{\text{apx},j}^{(k - 1)},1 \le j \le N_c\right.} \right\},
\end{equation}
where, $\mathbf{a}_{\text{apx},j}$ is the $j$-th row of $\mathbf{A}_{\text{apx}}$ and $b^{(k - 1)}_{\text{apx},j}$ is the $j$-th component of $\mathbf b^{(k - 1)}_{\text{apx}}$. Then, according to \cite{wangAggregateFlexibilityVirtual2021}, the boundary parameter $\mathbf b_{\text{apx}}$ is modified to make $\mathbf{P}^{(k)}_0$ lie on the boundary of $\Psi^{(k)}_{\text{apx}}$ by solving
\begin{subequations}\label{eq:mod_b}
  \begin{align}
      &\{b_{\text{apx},j}^{(k)}\}_{j\in \mathcal J^{(k)}}=\arg\max \sum\nolimits_{j \in {{{\mathcal J}}^{(k)}}} {{b_{{\text{apx}},j}}} \label{eq:mod_b_obj}\\
      &\text{s.t.  }{b_{{\text{apx}},j}} \le b_{{\text{apx}},j}^{(k - 1)},\forall j \in {{{\mathcal J}}^{(k)}},\label{eq:mod_b_1}\\
      &{{\mathbf{a}}_{{\text{apx}},j}}{{\mathbf{P}}_1} \le {b_{{\text{apx}},j}},\forall j \in {{{\mathcal J}}^{(k)}},\label{eq:mod_b_2}\\
      &{{\mathbf{a}}_{{\text{apx}},j}}{\mathbf{P}}_0^{(k)} \ge {b_{{\text{apx}},j}} - M(1 - {z_j}),\forall j \in {{{\mathcal J}}^{(k)}},\label{eq:mod_b_3}\\
      &z_j\in \mathbb B,\forall j \in {{{\mathcal J}}^{(k)}},\label{eq:mod_b_4}\\
      &\sum\nolimits_{j \in {{{\mathcal J}}^{(k)}}} {{z_j}}  \ge T.\label{eq:mod_b_5}
    \end{align}
\end{subequations}
The objective \eqref{eq:mod_b_obj} is to maximize the size of $\Psi^{(k)}_{\text{apx}}$. Constraints \eqref{eq:mod_b_1} and \eqref{eq:mod_b_2} guarantee $\Psi_{\text{apx}}^{(k)} \subseteq \Psi_{\text{apx}}^{(k-1)} $ and $\Psi_{\text{apx}}^{(k)} \neq \varnothing$, respectively. Constraints \eqref{eq:mod_b_3}-\eqref{eq:mod_b_5} together ensure that at least $T$ constraints in $\Psi_{\text{apx}}^{(k)}$ are active at point ${\mathbf{P}}_0^{(k)}$.

\begin{figure}[!t]
  \centering
  \includegraphics[width=3.4in]{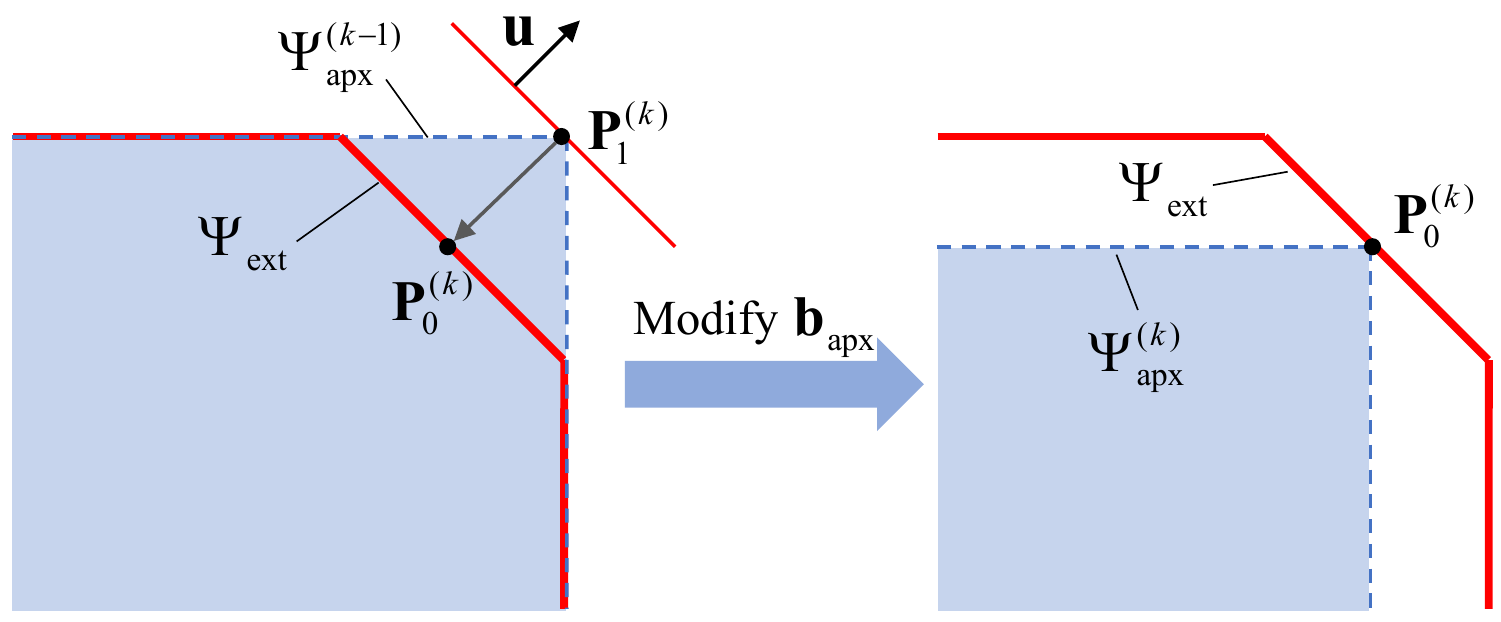}
  \caption{The calculation process for one modification on ${\mathbf{b}_{\text{apx}}}$.}
  \label{fig:illustration}
  \vspace{-0.15in}
\end{figure}

Fig. \ref{fig:illustration} illustrates the calculation process for one modification on ${\mathbf{b}_{\text{apx}}}$: 1) find direction $\mathbf u$ with the largest gap between $\Psi_{\text{apx}}^{(k-1)} $ and $\Psi_{\text{ext}}$ and get ${\mathbf{P}}_1^{(k)}$ correspondingly; 2) find ${\mathbf{P}}_0^{(k)}$ in $\Psi_{\text{ext}}$ that minimizes the distance between ${\mathbf{P}}_0^{(k)}$ and ${\mathbf{P}}_1^{(k)}$; 3) modify ${\mathbf{b}_{\text{apx}}}^{(k-1)}$ to ${\mathbf{b}_{\text{apx}}}^{(k)}$ to put ${\mathbf{P}}_0^{(k)}$ onto the boundary of $\Psi_{\text{apx}}^{(k)} $. 

The above discussions on the parameter calculation are summarized into the following algorithm:
\begin{itemize}[\IEEEusemathlabelsep\IEEEsetlabelwidth{Steps}]
  \item[Step 1:] Initialize the $N_c$ components of $\mathbf{b}_{\text{apx}}$ one by one via the parameter calculation formula in \cite{wenAggregateFeasibleRegion2022}, set a convergence tolerance $\epsilon>0$, $k=1$;
  \item[Step 2:] Solve the positive-direction problem \eqref{eq:trans_to_final} to get ${\mathbf{P}}_1^{(k)}$ and $d^{(k)}_{\text{pos}}$. If $d^{(k)}_{\text{pos}}>0$, go to Step 3, otherwise go to Step 4;
  \item[Step 3:] Solve problem \eqref{eq:find_P0} to get ${\mathbf{P}}_1^{(k)}$ and update ${\mathbf{b}^{(k)}_{\text{apx}}}$ according to \eqref{eq:mod_b}, $k\leftarrow k+1$;
  \item[Step 4:] Solve the negative-direction problem \eqref{eq:trans_to_final_neg} to get ${\mathbf{P}}_1^{(k)}$ and $d^{(k)}_{\text{neg}}$. If $d^{(k)}_{\text{neg}}>0$, go to Step 5, otherwise go to Step 6;
  \item[Step 5:] Do the same as Step 3;
  \item[Step 6:] If condition \eqref{eq:convergence} holds, finish iteration and output ${\mathbf{b}^{(k)}_{\text{apx}}}$, otherwise go to Step 2.
\end{itemize}
\begin{equation}\label{eq:convergence}
  \max\{d^{(k)}_{\text{pos}}, d^{(k)}_{\text{neg}}\}<\epsilon
\end{equation}

A noteworthy detail is that the iteration alternates between the positive and negative directions, rather than solely taking the direction with a larger gap, which reduces the number of optimization problems that need to be solved.
\subsubsection{Convergence Discussion}\;

In Step 6 of the above algorithm, the convergence condition \eqref{eq:convergence} indicates that the gap between $\Psi_{\text{apx}}^{(k)} $ and $\Psi_{\text{ext}}$ is sufficiently small in both positive and negative directions. This implies that $\Psi_{\text{apx}}^{(k)} \subseteq \Psi_{\text{ext}}$, which is precisely the desired outcome.

We have the following \textbf{Proposition} on the convergence:
\newtheorem{proposition}{Proposition}
\begin{proposition}
  \label{clm:converge}
  After a finite number of iterations, condition \eqref{eq:convergence} can always be reached.
\end{proposition}%

\emph{Proof:} The optimization problem \eqref{eq:mod_b} guarantees that $\Psi_{\text{apx}}^{(k)} \subseteq \Psi_{\text{apx}}^{(k-1)} $, which directly gives $d^{(k)}_{\text{pos}}\leq d^{(k-1)}_{\text{pos}}$ and $d^{(k)}_{\text{neg}}\leq d^{(k-1)}_{\text{neg}}$ by definition. Therefore, each modification on ${\mathbf{b}_{\text{apx}}}$ reduces the gap between $\Psi_{\text{apx}}^{(k)} $ and $\Psi_{\text{ext}}$ to zero in the corresponding direction $\mathbf u$ without increasing the gap of other directions. In the extreme case, all $\mathbf u \in \mathbb B^T$ are traversed, then $d^{(k)}_{\text{pos}}\leq 0 $ and $d^{(k)}_{\text{neg}}\leq 0$ hold. So condition \eqref{eq:convergence} must be reached in a finite number of iterations.$\hfill\blacksquare$ 

\section{Incorporating Distribution Network Constraints into Inner Approximation}

As adjustments of DERs can impact the distribution network, it is necessary to consider the distribution network constraints when aggregating power flexibility to the transmission-distribution interface. We employ the well-known LinDistFlow model to describe the distribution network constraints since it is computationally tractable and relatively accurate \cite{baranOptimalSizingCapacitors1989,hassanOptimalLoadEnsemble2019}. See Appendix \ref{sect:LinDistFlow} for details.

Based on our findings in \cite{wenAggregateTemporallyCoupled2022}, the expression for the aggregated power flexibility at the substation level, considering linear distribution network constraints, can still be accurately formulated as Equation \eqref{eq:EAM_vec} (with different $\underline {\phi} _{{\mathbf u}}$-s and $\overline {\phi} _{{\mathbf u}}$-s). Therefore, in theory, one can also construct matrices $\mathbf C$, $\mathbf D$, and vector $\mathbf f$ according to the flexible ranges of all DERs in the distribution network and network constraints, and apply the aforementioned algorithm to compute the inner approximated flexibility at the substation level. However, this approach may face scalability barriers when dealing with a large amount of DERs and a large-scale distribution network. In practice, on the other hand, the ADN operator may not have access to information at the device level since DERs are typically managed by a third-party aggregator.

To address these computational complexity and data privacy concerns, we propose a \textbf{double inner approximation framework} to model the power flexibility of the substation. At the first level, the DER aggregator builds the inner approximated flexibility model of its node in the distribution network using the algorithm described in Section \ref{sect:innerapprox}. At the second level, the distribution network operator builds an inner approximated flexibility model for the substation, considering all the aggregators' power flexibility and network constraints. The algorithm for the second level is similar to that for the first level, with some differences in the details. 

Specifically, for the calculation at the substation level, ${\Psi_{{\text{ext}}}}$ is represented as $\left\{ {{\mathbf{P}}_0} | {{\mathbf{C'x'}} - {{\mathbf{P}}_0} = {\mathbf{0}}, {\mathbf{E'x'}} = {\mathbf{g'}}, {\mathbf{D'x'}} \le {\mathbf{f'}}}  \right\}$, where $\mathbf{x}'$ contains all variables in the distribution network; ${\mathbf{C'x'}} - {{\mathbf{P}}_0} = {\mathbf{0}}$ corresponds to the active power balance equation \eqref{eq:LinDistFlow:1} at the substation; ${\mathbf{E'x'}} = {\mathbf{g'}}$ corresponds to all the other equality constraints at the distribution level, i.e., \eqref{eq:LinDistFlow:2}-\eqref{eq:LinDistFlow:6}; and ${\mathbf{D'x'}} \le {\mathbf{f'}}$ corresponds to voltagle limits \eqref{eq:LinDistFlow:7} and the inner approximated power flexibility ranges of DER aggregators. Correspondingly, the initialization of $\mathbf{b}_{\text{apx}}$ should be changed: each component of vector $\mathbf{b}^{(0)}_{\text{apx}}$ is calculated by
\begin{equation}
  \begin{array}{c}
  b^{(0)}_{\text{apx},j} = \mathop {\max } \mathbf{a}_{\text{apx},j}{{\mathbf{P}}_0} \\
  {\text{s.t.  }}{{\mathbf{C'x'}} - {{\mathbf{P}}_0} = {\mathbf{0}}, {\mathbf{E'x'}} = {\mathbf{g'}}, {\mathbf{D'x'}} \le {\mathbf{f'}}},
  \end{array}
  \nonumber
\end{equation}
where, $\mathbf{a}_{\text{apx},j}$ represents the $j$-th row in the coefficient matrix $\mathbf{A}_{\text{apx}}$. Besides, the original optimization problem \eqref{eq:trans_to_final} that calculates the maximum gap between the approximation model and the EAM in the positive directions should be changed to
\begin{subequations}\label{eq:cal_gap_substation}
  \begin{align}
    &\min  {{\mathbf{1}}^{\top}}\mathbf y + {{\boldsymbol{\lambda }'}^{\top}}{\mathbf{f'}} + {{\boldsymbol{\mu}'}^{\top}}{\mathbf{g'}}  \label{eq:cal_gap_substation_obj}\\
    &\text{s.t.  }{\mathbf{u}} \in {\mathbb B ^T},{{\mathbf{P}}_1},{\boldsymbol{\omega }'}, \mathbf y \in {\mathbb{R}^T},{\boldsymbol{\lambda }'} \in {\mathbb{R}^{m_1}},{\boldsymbol{\mu }'} \in {\mathbb{R}^{m_2}},\nonumber \\
    &{{\mathbf{A}}_{{\text{apx}}}}{{\mathbf{P}}_1} \le {{\mathbf{b}}^{(k-1)}_{{\text{apx}}}},\label{eq:cal_gap_substation1}\\
    &{\mathbf{u}} + {\boldsymbol{\omega }}' = \mathbf{0},\label{eq:cal_gap_substation2}\\
    &{{\mathbf{C'}}^{\top}}{\boldsymbol{\omega }}' + {{\mathbf{D'}}^{\top}}{\boldsymbol{\lambda }}' + {{\mathbf{E'}}^{\top}}{\boldsymbol{\mu }'} = \mathbf{0},\label{eq:cal_gap_substation3}\\
    &{\boldsymbol{\lambda }'} \ge \mathbf{0},{\boldsymbol{\mu}'} \ge \mathbf{0},\label{eq:cal_gap_substation4} \\
    &-M\mathbf u\leq \mathbf y \leq M\mathbf u\label{eq:cal_gap_substation5},\\
    & - M(\mathbf 1 - {\mathbf{u}}) \le \mathbf y +\mathbf P_1 \le M(\mathbf 1 - {\mathbf{u}}),\label{eq:cal_gap_substation6}
  \end{align}
  \end{subequations}
where ${\boldsymbol{\omega }}'$, ${\boldsymbol{\lambda}}'$, and ${\boldsymbol{\mu}}'$ are dual variables of ${\mathbf{C'x'}} - {{\mathbf{P}}_0} = {\mathbf{0}}$, ${\mathbf{D'x'}} \le {\mathbf{f'}}$, and ${\mathbf{E'x'}} = {\mathbf{g'}}$, respectively; $m_1$ and $m_2$ represent the numbers of lines in ${\mathbf{D'x'}} \le {\mathbf{f'}}$ and ${\mathbf{E'x'}} = {\mathbf{g'}}$, respectively. Similar modifications should also be made to the optimization problems \eqref{eq:trans_to_final_neg} and \eqref{eq:find_P0} due to the change in ${\Psi _{{\text{ext}}}}$, which are not included here for the sake of simplicity.

By implementing the proposed double inner approximation framework, it is ensured that the power profile of the substation can be disaggregated into each aggregator, and that the total power profile of each aggregator can be disaggregated into each DER. However, a natural concern is that this framework may sacrifice accuracy and result in excessive conservatism. We note that this is a necessary trade-off for guaranteeing computational tractability and data privacy. On the other hand, this shortcoming can be mitigated by using higher precision prototypes at both levels, such as the energy-change boundary prototype. The efficiency of the proposed framework will be verified in the case studies.

\section{Case Studies}

We first test the inner approximated aggregation at the aggregator level in Subsection \ref{sect:test_no_DN}. Next, the flexibility aggregation in an ADN interconnected with multiple aggregators is tested in Subsection \ref{sect:test_DN}. Finally, we solve an SCUC problem that considers the inner approximated flexibility of ADNs in Subsection \ref{sect:test_SCUC} to verify the applicability of the proposed method. 

In the simulations, the convergence tolerance $\epsilon$ is set to $10^{-4}$. All simulations were performed on a desktop with a 16-core Intel i7-10700 processor and 32 GB RAM, programmed in MATLAB and solved by Cplex.

\subsection{Test the Inner Approximation for Aggregating Flexibility at the Aggregator Level}\label{sect:test_no_DN}

Consider the aggregation of the power flexibility from 50 EVs. Each EV has a rated charging power of 7 kW, a battery capacity of 50 kWh, and a charging efficiency of 0.95. The arrival and departure times and states-of-charge are generated using Monte Carlo simulation. The proposed inner approximation method with two different prototypes: the power-energy boundary and energy-change boundary prototypes, is tested. 

\begin{figure}[!t]
  \centering
  \includegraphics[width=3.4in]{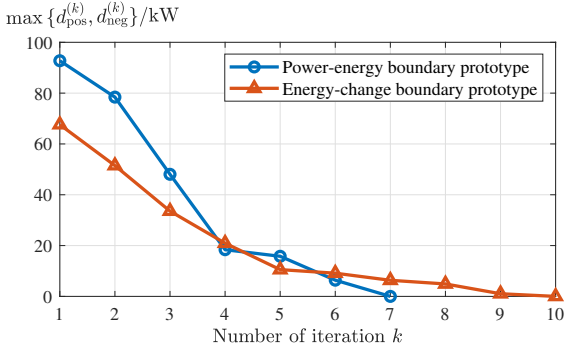}
  \caption{Changes in $\max{\{d_{\text{pos}}^{(k)},d_{\text{neg}}^{(k)}\}}$ during the iteration of the proposed methods with two prototypes.}
  \label{fig:iteration_process}
\end{figure}

\begin{figure}[t!]
  \begin{centering}
  \subfigure[Power boundaries]{\includegraphics[width=3.4in]{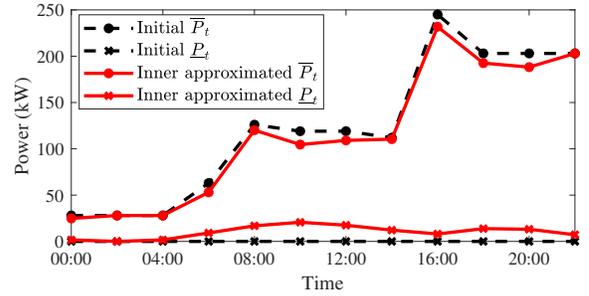}}
  \subfigure[Energy boundaries]{\includegraphics[width=3.4in]{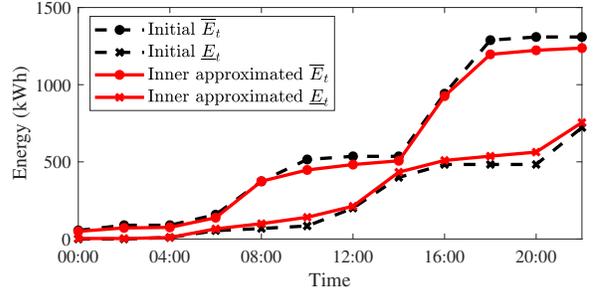}}
  \end{centering}
  \caption{Change in power and energy boundaries made by the proposed inner approximation method.}
  \label{fig:comp_PE}
\end{figure}

We first set $T=12$ and run the inner approximation algorithm. The iteration process under two prototypes are illustrated in Fig. \ref{fig:iteration_process}. The maximum gap $\max{\{d_{\text{pos}}^{(k)},d_{\text{neg}}^{(k)}\}}$ reduces gradually as expected. Convergence is attained within 7 iterations for the power-energy boundary prototype and within 10 iterations for the energy-change boundary prototype. Taking the power-energy boundary prototype as an example, Fig. \ref{fig:comp_PE} illustrates the changes in these boundary parameters caused by the inner approximation. Both the power and energy boundaries shrink slightly inward from the initial values, leading to a change in the feasible region determined by the power-energy boundary model from containing EAM to being contained by EAM.

The proposed method is further compared with the inner approximation methods in \cite{chenAggregatePowerFlexibility2020} and \cite{wangAggregateFlexibilityVirtual2021} in terms of both accuracy and computational efficiency. We measure the accuracy of the inner approximation model via the following steps. First, generate $N_{\hat{\mathbf u}}$ different directions $\hat{\mathbf u}$ randomly from $\mathbb B^{T}$. For each $\hat{\mathbf u}$, calculate the maximum length, denoted by $L_{\text{apx},\hat{\mathbf u}}$, that the approximation model can extend in direction $\hat{\mathbf u}$ by solving linear program \eqref{eq:def_L_apx}. The length $L_{\text{apx},\hat{\mathbf u}}$ must not exceed the maximum length $L_{\text{ext},\hat{\mathbf u}}$ (calculated by \eqref{eq:def_L_ext}), that the EAM can extend in direction $\hat{\mathbf u}$, because inner approximation models must be inside the EAM. Finally, we calculate the geometrical average of the rate ${L_{\text{apx},\hat{\mathbf u}}}/{L_{\text{ext},\hat{\mathbf u}}}$ over all $N_{\hat{\mathbf u}}$ $\hat{\mathbf u}$-s, namely \textbf{relative size}. A larger relative size indicates a more accurate, i.e. less conservative model.
\begin{align}
  {L_{{\text{apx}},{\hat{\mathbf u}}}} &\triangleq \mathop {\max }\limits_{{\mathbf{P}_+},{\mathbf{P}_-} \in {\Psi _{{\text{apx}}}}} {\frac{\hat{\mathbf u}^{\top}}{\left\| {\hat{\mathbf u}}\right\|}}\left({\mathbf{P}_+}-{\mathbf{P}_-}\right) \label{eq:def_L_apx}\\
  {L_{{\text{ext}},{\hat{\mathbf u}}}} &\triangleq \mathop {\max }\limits_{{\mathbf{P}_+},{\mathbf{P}_-} \in {\Psi _{{\text{ext}}}}} {\frac{\hat{\mathbf u}^{\top}}{\left\| {\hat{\mathbf u}}\right\|}}\left({\mathbf{P}_+}-{\mathbf{P}_-}\right) \label{eq:def_L_ext}
\end{align}

Table \ref{tab:relative_size} lists the relative sizes of different models with $N_{\hat{\mathbf u}}=50$. It can be observed that the proposed method with the energy-change boundary prototype has the largest relative size among all the approximation methods, indicating its high accuracy (low conservatism). Using the same power-energy boundary prototype, the relative size of the proposed method is similar to that of \cite{wangAggregateFlexibilityVirtual2021}. The conservatism of the method in \cite{chenAggregatePowerFlexibility2020} is the highest because its prototype, i.e., the power boundary prototype, is far from accurate.

\begin{table}[!htbp]
  \captionsetup{justification=centering, labelsep=newline}
  \caption{\textsc{Relative Sizes Resulting from Different Inner Approximation Methods}}
  \centering
   \begin{tabular}{ccc}
    \toprule
    Method & Geometric prototype & Relative size\\
    \hline
    \cite{chenAggregatePowerFlexibility2020} & Power boundary & 0.2291 \\
    \hline
    \cite{wangAggregateFlexibilityVirtual2021} & Power-energy boundary & 0.7766    \\
    \hline
    \multirow{2}{*}{$\begin{array}{c}\textbf{Proposed} \\ \textbf{method}\end{array}$} & Power-energy boundary & 0.8239 \\
    \cline{2-3}
     & Energy-change boundary & 0.9302   \\
     \hline
  \end{tabular}
  \label{tab:relative_size}
 \end{table}


At the end of this subsection, we present the result of a scalability test. When $T=12$, the computation times of the proposed method under the power-energy boundary prototype and the energy-change boundary prototype are 1.910 s and 3.528 s, respectively. These times are significantly lower than the 35.36 s required by the method in \cite{wangAggregateFlexibilityVirtual2021}. When $T=24$, convergence is achieved within 46.41 s under the power-energy boundary prototype and 138.9 s under the energy-change boundary prototype. In contrast, the method in \cite{wangAggregateFlexibilityVirtual2021} does not converge within 1000 s. Therefore, we can conclude that the proposed method yields a substantial improvement in computational efficiency.

 \subsection{Test the Inner Approximation for Aggregating Flexibility at the Substation Level}\label{sect:test_DN}
 In this subsection, we set $T=24$ and implement the inner approximated flexibility aggregation for the ADN based on a modified IEEE-33 bus distribution network. The base loads in the 33-bus network are multiplied by three to match the load level of later SCUC. The branch impedances are divided by three to maintain the nodal voltages similar to the original network. The voltage limits are set to 1.10 p.u. and 0.90 p.u. Half of the nodes in the distribution network are connected to a DER aggregator, which consists of 40 EVs, 1 PV of 20 kW installed capacity, and 1 DESS. The maximum output profile data of the PVs are from a southern province in China. The DESSs have a rated charging and discharging power of 10 kW, a capacity of 80 kWh, and an efficiency stochastically generated between 0.92 and 0.97. The DESSs' energy is required to be restored to the initial value at the end of the time horizon.

In the proposed double inner approximation framework, the prototypes at the aggregator and substation levels can be different. We run the proposed method based on four different prototype combinations:
\begin{itemize}[\IEEEusemathlabelsep\IEEEsetlabelwidth{Model}]
  \item[Model 1:] Power-energy boundary at both levels;
  \item[Model 2:] Power-energy boundary at the aggregator level and energy-change boundary at the substation level;
  \item[Model 3:] Energy-change boundary at the aggregator level and power-energy boundary at the substation level;
  \item[Model 4:] Energy-change boundary at both levels.
\end{itemize}

For comparison, we also implement the approximation model in \cite{chenAggregatePowerFlexibility2020}. The model in \cite{wangAggregateFlexibilityVirtual2021} is not included this time because it has already encountered complexity barriers at the aggregator level. The accuracy of inner approximation models is also measured by the relative size calculated by the same method as in Subsection \ref{sect:test_no_DN}.

Table \ref{tab:performance_DN} lists each model's relative size, computation time, and number of constraints at the substation level. Although the model in \cite{chenAggregatePowerFlexibility2020} has the lowest computation time, its accuracy is much lower than the four models generated by the proposed method. The accuracy of Models 3 and 4 is higher than that of Models 1 and 2, and correspondingly, the computation time is also higher. On the other hand, the difference in accuracy between models 1 and 2, as well as that between models 3 and 4, is not significant. This phenomenon suggests that using a higher-accuracy energy-change boundary model at the substation level has a less pronounced effect than at the aggregator level. Considering that the flexibility model at the substation level is to be used for subsequent transmission network scheduling, its number of constraints should not be too large. Therefore, the most recommended combination is model 3, i.e., energy-change boundary at the aggregator level and power-energy boundary at the substation level.
\begin{table}[!htbp]
  \centering
  \captionsetup{justification=centering, labelsep=newline}
  \caption{\textsc{Comparison of Different Flexibility Aggregation Models at the Substation Level}}
   \begin{tabular}{cccc}
    \toprule
    Model & $\begin{array}{c}\text{Relative size}\\ \text{(Accuracy)}\end{array}$ & $\begin{array}{c}\text{Computation}\\ \text{time (s)}\end{array}$ & $\begin{array}{c}\text{Number of}\\ \text{constraints}\end{array}$\\
    \midrule
    \cite{chenAggregatePowerFlexibility2020}& 0.2158 & 3.324 & 48\\
    Model 1& 0.3995 & 50.43 & 94\\
    Model 2& 0.4060 & 66.48 & 600\\
    Model 3& 0.5464 & 657.8 & 94\\
    Model 4& 0.5641 & 925.1 & 600\\
   \bottomrule
  \end{tabular}
  \label{tab:performance_DN}
 \end{table}

 \subsection{Test SCUC with Aggregated Flexibility of ADNs}\label{sect:test_SCUC}

 SCUC is a widely-used decision-making tool in power systems that schedules generators to supply load demand while minimizing the total cost and satisfying the system security constraints at the transmission level. ADNs hold considerable power flexibility when they aggregate a large number of DERs, thereby can be utilized in SCUC to enhance the coordination between transmission and distribution networks \cite{kargarianSystemSystemsBased2014}. The proposed inner approximation method outputs a realizable power range of ADNs, which is very suitable for modeling ADNs in the SCUC problem. This subsection tests the SCUC using an IEEE 30-bus transmission network interconnected with 13 IEEE 33-bus ADNs. Overall, the SCUC in question is a Mixed-Integer Quadratic Programming problem. It includes DC power flow equations of the transmission system, modeling of traditional generators with reserve for uncertainty, scenario-based uncertainty modeling for wind generation, and modeling for power flexibility of ADNs. Detailed formulations can be found in Appendix \ref{sect:SCUC}.

 In the SCUC problem, $T$ is set to 24 \cite{tangReserveModelEnergy2021,gutierrez-alcarazLargeScalePreventiveSecurityConstrained2022}. Model 3, as recommended in the previous subsection, is used to model the power flexibility of ADNs. The generation cost parameters of traditional generators are the same as in \cite{wenAggregateTemporallyCoupled2022}. The cost for scheduling reserve is fixed at \$1/MWh \cite{pozoUnitCommitmentIdeal2014}. Five wind farms are put in the transmission network, each with an installed capacity of 10 MW. The number of scenarios for wind generation is set to 10, and the data comes from \cite{pinson2013wind}. 

 We solve the SCUC problem with ADNs and compare the results to the case where distribution networks operate at the baseline\footnote{The baseline power is calculated by minimizing the total distance of the power profile from each boundary in the flexibility model.}. The total cost of the former is \$116,617, while the latter is \$118,506, resulting in a cost reduction of \$1,889. This cost reduction is relatively low due to the small ratio of flexible resources to fixed load in the system, which can be more significant with an increase in the number of DERs.

\begin{figure}[!htbp]
  \centering
  \includegraphics[width=3.4in]{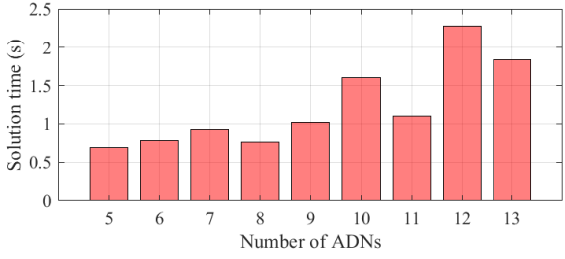}
  \caption{Solution time of the SCUC problem with different numbers of ADNs.}
  \label{fig:sol_time}
\end{figure}

 Finally, we run the SCUC program for the 30-bus transmission network with 5 to 13 ADNs. The solution time is shown in Fig. \ref{fig:sol_time}, from which we can see that the number of ADNs will not significantly affect the computation time. This is because the proposed flexibility aggregation model is very concise: it includes only $4T-2$ constraints for each ADN from the perspective of the transmission network operator. Hence, the proposed flexibility modeling for ADNs exhibits high scalability in the SCUC problem, making it a desirable choice for practical implementation.

\section{Conclusions}
This paper proposes an improved inner approximation method for the power flexibility of ADNs, which enables concise and reliable modeling of ADNs in upper-level decision-making. The main idea of the proposed method is to leverage the property of the EAM in the inner approximation process. Specifically, we use the energy-change boundary model derived from a subset of constraints in the EAM to define the prototype of the approximation model, which substantially enhances accuracy (or, in other words, reduces conservatism). Also, we take advantage of the fact that the coefficient vectors in EAM are binary vectors in designing a parameter calculation algorithm, significantly improving the computational efficiency. 

Furthermore, we propose a double inner approximation framework for aggregating the flexibility to the substation level while considering network constraints in the ADN. Although this framework sacrifices accuracy relative to the single-level inner approximation, it is more practical because it ensures computational tractability for a large scale of DERs and data privacy for end-users. Case studies also show that the double inner approximation does not lead to too much accuracy loss. Finally, the proposed inner approximated flexibility model of ADNs is applied to the SCUC problem in numerical simulation, revealing the practicality of the proposed method.

\appendices
\section{The LinDistFlow Formulation}\label{sect:LinDistFlow}
\setcounter{equation}{0}
\renewcommand\theequation{A.\arabic{equation}}
The LinDistFlow model of the distribution network is formulated as:
\begin{equation}
 {P_{0,t}} = \sum\nolimits_{j \in 0} {{P_{0j,t}}}+ P_{0,t}^{{\text{fix}}} + P_{0,t}^{{\text{flex}}},\forall t\in [T],\label{eq:LinDistFlow:1}
\end{equation}
\begin{equation}
  \sum\nolimits_{j \in i} {{P_{ij,t}}}  + P_{i,t}^{{\text{fix}}} + P_{i,t}^{{\text{flex}}} = 0,\forall i \in {{{\mathcal I}}^ + },t\in [T],\label{eq:LinDistFlow:2}
 \end{equation}
 \begin{equation}
  \sum\nolimits_{j \in i} {{Q_{ij,t}}}  + Q_{i,t}^{{\text{fix}}} + Q_{i,t}^{{\text{flex}}} = 0,\forall i \in {{{\mathcal I}}},t\in [T],\label{eq:LinDistFlow:3}
 \end{equation}
 \begin{equation}
  {V_{i,t}} - {V_{j,t}} = 2({r_{ij}}{P_{ij,t}} + {x_{ij}}{Q_{ij,t}}),\forall ij \in {{\mathcal L}},t\in [T],\label{eq:LinDistFlow:4}
 \end{equation}
 \begin{equation}
  Q_{i,t}^{{\text{flex}}} = P_{i,t}^{{\text{flex}}}\tan {\gamma _i},\forall i \in {{{\mathcal I}} },t\in [T],\label{eq:LinDistFlow:5}
 \end{equation}
 \begin{equation}
  {V_{0,t}} = 1,\forall t\in [T],\label{eq:LinDistFlow:6}
 \end{equation}
 \begin{equation}
  {\underline V _i} \le {V_{i,t}} \le {\overline V _i},\forall i \in {{{\mathcal I}}^ + },t\in [T],\label{eq:LinDistFlow:7}
 \end{equation}
 \begin{equation}
  P_{i,t}^{{\text{flex}}} = \sum\nolimits_{n \in {{{\mathcal N}}_i}} {{p_{n,t}}},\forall i \in {{{\mathcal I}}},t\in [T],\label{eq:LinDistFlow:8}
 \end{equation}
 \begin{equation}
  {\underline p _{n,t}} \le {p_{n,t}} \le {\overline p _{n,t}},\forall n \in {{\mathcal N}},t\in [T] ,\label{eq:LinDistFlow:9}
 \end{equation}
 \begin{equation}
  {\underline e _{n,t}} \le \sum\nolimits_{\tau  = 1}^t {{p_{n,\tau }}\Delta T}  \le {\overline e _{n,t}},\forall n \in {{\mathcal N}},t\in [T],\label{eq:LinDistFlow:10}   
 \end{equation}
  where, $i/\mathcal I$ denotes the indices/set of nodes in the distribution network, $0$ represents the substation (root node), and $\mathcal I^+\triangleq \mathcal I \backslash \{0\}$; $\mathcal N_i$ denotes the set of DERs located at node $i$, $\bigcup_{i\in \mathcal I}\mathcal N_i = \mathcal N$; $\mathcal L$ denotes the set of lines where $ij$ refers to the line between node $i$ and node $j$; $j\in i$ indicates that node $j$ is connected to node $i$; $P_{0,t}$ denotes the active power injection of the root node at time $t$; $P_{ij,t}/Q_{ij,t}$, $P^{\text{fix}}_{i,t}/Q^{\text{fix}}_{i,t}$ and $P^{\text{flex}}_{i,t}/Q^{\text{flex}}_{i,t}$ denote the active/reactive power flow in line $ij$, fixed load at node $i$, and flexible power at node $i$, respectively; $V_{i,t}$ is the square of voltage at node $i$; $r_{ij}$ and $x_{ij}$ denote the resistance and reactance of branch $ij$, respectively; $\gamma_i$ denotes the power factor angle of the DER aggregator at node $i$; $\overline V_{i,t}$ and $\underline V_{i,t}$ denote the square of upper and lower voltage limit of node $i$, respectively; $p_{n,t}$ is the power of DER $n$; and $\underline p_{n,t}$, $\overline p_{n,t}$, $\underline e_{n,t}$, and $\overline e_{n,t}$ are the power and energy boundaries of DER $n$, respectively. 

  Equation \eqref{eq:LinDistFlow:1} defines the power at the substation. Constraints \eqref{eq:LinDistFlow:2}-\eqref{eq:LinDistFlow:4} are the LinDistFlow equations. Constraint \eqref{eq:LinDistFlow:5} represents the relationship between the active and reactive power of DER aggregators while assuming a constant power factor angle. Equation \eqref{eq:LinDistFlow:6} sets the voltage of the root node to 1 p.u. and \eqref{eq:LinDistFlow:7} restricts the voltage of other nodes in the distribution network. Equation \eqref{eq:LinDistFlow:8} defines the active power of the aggregator as the sum of the active power of each DER under its control. And constraints \eqref{eq:LinDistFlow:9} and \eqref{eq:LinDistFlow:10} represent the same power and energy boundaries of each DER as described in \eqref{eq:powerenergybound}.

  \section{The SCUC Formulation}\label{sect:SCUC}
\setcounter{equation}{0}
\renewcommand\theequation{B.\arabic{equation}}
The SCUC problem solved in the case studies is formulated mainly based on reference \cite{tangReserveModelEnergy2021}. Uncertainty of wind power is considered using the scenario-based stochastic programming. We do not create any specific marker to distinguish the variables and parameters in the SCUC from those in the LinDistFlow model in Appendix A, which will not lead to confusion since the computation of SCUC and LinDistFlow are separated. The SCUC problem is formulated as follows:
\begin{equation}
  \min \sum\limits_{t \in [T]} {\left[ \begin{array}{l}
    \sum\limits_{s \in {{\mathcal S}}} {pro{b_s} \cdot \sum\limits_{g \in {{\mathcal G}}} {\left( {{a_g}P_{g,s,t}^2 + {b_g}{P_{g,s,t}} + {c_g}} \right)} } \\
      + \sum\limits_{g \in {{\mathcal G}}} {\left( \begin{array}{l}
    C_g^{{\text{SU}}}{y_{g,t}} + C_g^{{\text{SD}}}{z_{g,t}}\\
      + C_g^{\text{U}}R_{g,t}^{\text{U}} + C_g^{\text{D}}R_{g,t}^{\text{D}}
    \end{array} \right)} 
    \end{array} \right]} \label{eq:SCUC:obj}
\end{equation}
s.t.
\begin{equation}
  {P_{ij,s,t}} = \frac{{{\theta _{i,s,t}} - {\theta _{j,s,t}}}}{{{X_{ij}}}},\forall ij \in {{\mathcal L}},s \in {{\mathcal S}},t \in [T],\label{eq:SCUC:1}
\end{equation}
\begin{equation}
  {\theta _{0,s,t}} = 0,\forall s \in {{\mathcal S}},t \in [T],\label{eq:SCUC:2}
\end{equation}
\begin{equation}
  \begin{array}{l}
    \sum\limits_{j \in i} {{P_{ij,s,t}}}  + \sum\limits_{d \in {{{\mathcal D}}_i}} {{P_{d,s,t}}}  + P_{i,t}^{\text{F}} - \sum\limits_{g \in {{{\mathcal G}}_i}} {{P_{g,s,t}}} \\
    \;\;\;\;\;\;\;\;\;\;\;\;- \sum\limits_{w \in {{{\mathcal W}}_i}} {{P_{w,s,t}}}  = 0,\forall i \in {{\mathcal I}},s \in {{\mathcal S}},t \in [T],
    \end{array}\label{eq:SCUC:3}
\end{equation}
\begin{equation}
  - P_{ij}^{\text{U}} \le {P_{ij,s,t}} \le P_{ij}^{\text{U}},\forall ij \in {{\mathcal L}},s \in {{\mathcal S}},t \in [T],\label{eq:SCUC:4}
\end{equation}
\begin{equation}
  0 \le {P_{w,s,t}} \le P_{w,s,t}^{\text{U}},\forall w \in {{\mathcal W}},s \in {{\mathcal S}},t \in [T], \label{eq:SCUC:5}
\end{equation}
\begin{equation}
  {u_{g,t}},{y_{g,t}},{z_{g,t}} \in \mathbb B ,\forall g \in {{\mathcal G}},\forall t \in [T],\label{eq:SCUC:6}
\end{equation}
\begin{equation}
  \sum\limits_{\tau  = t}^{t + T_g^{{\text{on}}} - 1} {{u_{g,\tau }}}  \ge T_g^{{\text{on}}}  {y_{g,t}},\forall g \in {{\mathcal G}},\forall t \in [T - T_g^{{\text{on}}} + 1],\label{eq:SCUC:7}
\end{equation}
\begin{equation}
  \sum\limits_{\tau  = t}^{t + T_g^{{\text{off}}} - 1} {\left( {1 - {u_{g,\tau }}} \right)}  \ge T_g^{{\text{off}}}  {z_{g,t}},\forall g \in {{\mathcal G}},\forall t \in [T - T_g^{{\text{off}}} + 1],\label{eq:SCUC:8}
\end{equation}
\begin{equation}
  {u_{g,t+1}} - {u_{g,t}} = {y_{g,t+1}} - {z_{g,t+1}},\forall g \in {{\mathcal G}},\forall t \in [T-1],\label{eq:SCUC:9}
\end{equation}
\begin{equation}
  {y_{g,t}} + {z_{g,t}} \le 1,\forall g \in {{\mathcal G}},\forall t \in [T],\label{eq:SCUC:10}
\end{equation}
\begin{equation}
  P_g^{\text{L}} {u_{g,t}} \le {P_{g,s,t}} \le P_g^{\text{U}} {u_{g,t}},\forall g \in {{\mathcal G}},s \in {{\mathcal S}},t \in [T],\label{eq:SCUC:11}
\end{equation}
\begin{equation}
  \begin{array}{l}
  - {r_g^{\text{D}} {u_{g,t}} - r_g^{{\text{SD}}} {z_{g,t+1}}} \le {P_{g,s,t+1}} - {P_{g,s,t}}\\
  \le r_g^{\text{U}} {u_{g,t}} + r_g^{{\text{SU}}} {y_{g,t+1}},\forall g \in {{\mathcal G}},s \in {{\mathcal S}},t\in[T-1],
 \end{array}\label{eq:SCUC:12}
\end{equation}
\begin{equation}
  - R_{g,t}^{\text{D}} \le {P_{g,s,t}} - {P_{g,0,t}} \le R_{g,t}^{\text{U}},\forall g \in {{\mathcal G}},s \in {{\mathcal S}^+},t \in [T],\label{eq:SCUC:13}
\end{equation}
\begin{equation}
  R_{g,t}^{\text{U}} \ge 0,R_{g,t}^{\text{D}} \ge 0,\forall g \in {{\mathcal G}},t \in [T],\label{eq:SCUC:14}
\end{equation}
\begin{equation}
  {\mathbf{A}_d}{{\mathbf{P}}_{d,s}} \le {{\mathbf{b}}_d},\forall d \in {{\mathcal D}},s \in {{\mathcal S}},\label{eq:SCUC:15}
\end{equation}
  where, $s/ \mathcal S$ denotes the indices/set of scenarios of wind generation, $0$ represents the base scenario and $\mathcal S^+ = \mathcal S\backslash \{0\}$; $i/\mathcal I$ denotes the indices/set of nodes in the transmission network, node $0$ represents the slack node; $\mathcal L$ denotes the set of lines and $ij$ refers to the line between node $i$ and node $j$; $g/\mathcal G$, $d/\mathcal D$, and $w/\mathcal W$ denote the indices/set of generators, ADNs, and wind farms in the transmission network, respectively, and $\mathcal G_i$, $\mathcal D_i$, and $\mathcal W_i$ denotes the set of those located at node $i$, respectively; $P_{ij,s,t}$, $P_{g,s,t}$, $P_{d,s,t}$, $P_{w,s,t}$, $\theta_{i,s,t}$ denote the active power flow in line $ij$, the active output power of generator $g$, the active load of ADN $d$, the active generation of wind farm $w$, and the phase angle at node $i$ in scenario $s$ and time $t$, respectively; $P^{\text{F}}_{i,t}$ denotes the fixed load at node $i$; $u_{g,t}$ denotes the ON/OFF status of generator $g$, while $y_{g,t}/z_{g,t}$ is the startup/shutdown indicator; $R_{g,t}^{\text{D}}.R_{g,t}^{\text{U}}$ denotes the up/down reserves of generator $g$; $\mathbf P_{d,s}$ is the $T\times 1$ vector composed by $P_{d,s,t},\forall t \in [T]$; $prob_{s}$ denote the probability of scenario $s$; $a_g$, $b_g$, and $c_g$ are the generation cost parameters of generator $g$, respectively; $C_g^{\text{SU}}$, $C_g^{\text{SD}}$, $C_g^{\text{U}}$, and $C_g^{\text{D}}$ denote the costs for startup, shutdown, up reserve, and down reserve of generator $g$, respectively; $X_{ij}$ is the impedance of line $ij$, and $P^{\text{U}}_{ij}$ is the transmission capacity of line $ij$; $P^{\text{U}}_{w,s,t}$ is the maximum generation profile of wind farm $w$ in scenario $s$; $T_g^{\text{on}}/T_g^{\text{off}}$ denotes the minimum ON/OFF time of generator $g$, and $P_g^{\text{L}}/P_g^{\text{U}}$ denotes the minimum/maximum output limit of generator $g$; $r_g^{\text{U}}/r_g^{\text{D}}$ is the ramp up/down rate of generator $g$, and $r_g^{\text{SU}}/r_g^{\text{SD}}$ is the startup/shutdown ramp rate of generator $g$; $\mathbf A_d$ and $\mathbf b_d$ denote the coefficient matrix and right-hand vector of the inner approximated flexibility model of ADN $d$, respectively.

  The objective \eqref{eq:SCUC:obj} consists of two part: the first part is the expected generation cost over all scenarios, while the second part contains the costs of startup, shutdown, up reserve, and down reserve of generators. The penalty for wind curtailment is not explicitly included, as maximizing the utilization of wind power is already implicitly incorporated in the objective. Besides, the cost of adjusting power within the feasible region of ADNs is assumed to be negligible and hence not included in the objective function. These factors can be easily added to the objective when required.

  Constraint \eqref{eq:SCUC:1} represents the DC power flow equation. Equation \eqref{eq:SCUC:2} fixes the phase angle of the slack bus to $0$, thereby providing a reference for the other nodes. Constraint \eqref{eq:SCUC:3} enforces nodal power balance. Constraints \eqref{eq:SCUC:4} and \eqref{eq:SCUC:5} respectively limit the line power flow and wind generation. Constraints \eqref{eq:SCUC:7} and \eqref{eq:SCUC:8} impose minimum ON and OFF time limits on generators, respectively. The startup and shutdown logic of generators is captured via the classic "3-bin" form in constraints \eqref{eq:SCUC:9} and \eqref{eq:SCUC:10}. Generators' minimum and maximum outputs are constrained by \eqref{eq:SCUC:11}, while their ramp up and down abilities are restricted by constraint \eqref{eq:SCUC:12}. Constraint \eqref{eq:SCUC:13} requires that the generators' reserves cover at least the maximum offset of their output power in each scenario relative to the base scenario. Non-negativity restrictions on reserves are imposed via constraint \eqref{eq:SCUC:14}. Finally, the inner approximated feasible range for power adjustment of ADNs is compactly expressed in \eqref{eq:SCUC:15}.

\small
\bibliographystyle{IEEEtran}

\bibliography{ref}


\end{document}